\documentclass[a4paper]{article}
\usepackage{graphicx}
\newcommand{\be}{\begin{equation}}
\newcommand{\ee}{\end{equation}}
\newcommand{\bd}{\begin{displaymath}}
\newcommand{\ed}{\end{displaymath}}
\newcommand{\bea}{\begin{eqnarray}}
\newcommand{\eea}{\end{eqnarray}}
\newcommand{\bi}{\begin{description}}
\newcommand{\ei}{\end{description}}
\newcommand{\bq}{\begin{quote}}
\newcommand{\eq}{\end{quote}}
\def\i{\item}
\def\fo{\footnote}

\begin{document}
\bibliographystyle{unsrt}

\author{Alexander~Unzicker\fo{Corresponding author:alexander.unzicker et lrz.uni-muenchen.de}\  and Daniel~Schmidle\\
        Pestalozzi-Gymnasium  M\"unchen\\[0.6ex]}
\title{A Quick and Dirty Approach to Verify the Pioneer Anomaly}
\maketitle

\begin{abstract}

We present source code for the computer algebra system Mathematica
that analyzes the motion of the Pioneer spacecraft using the
public available ephemeris data from JPL's website. Within 15
minutes, the reader can verify that the Pioneer anomalous
acceleration $a_p$ (1) exists in the order of magnitude of $c
H_0$, (2) is not due to mismodeling of gravitational attraction,
solar pressure or spacecraft attitude maneuvers. The simple code
of about 100 lines may easily be extended by the reader to include
further tests. Due to the limitations of our approach, we do not
know (1) whether the unknown raw data were correctly processed to
generate the trajectory files (2) how the apparent mismatch of
ephemerides before 1990 had occurred.

\end{abstract}

\section{Introduction}
The Pioneer anomaly $a_p \approx -8.7 \times 10^{-10} m s^{-2}$
\cite{And:98, And:01} has become one of the major challenges for
theoretical physicists, and further efforts to investigate its
origin are underway \cite{Nie, Nie:04, Tur:06, Nie:07}. The
approximate coincidence of $a_p$ with the speed of light divided
by the age of the universe has raised the question if this effect
is new physics or a general failure of Newton's law of gravitation
\cite{And:01, And:06, Lam:06, Unz:07}. A couple of months ago, one
of us heard two astrophysicists talking about the anomaly. One of
them, probably tired of hearing about new trouble besides DM and
DE, concluded: `well, I still don't believe it...'.

Though the analysis of the Pioneer data was done by two independent groups
and published in peer-reviewed journals, it is reality that convincing
scientists needs time. Here we do not present any new
results that help to understand the anomaly, and our approach cannot compete
with the detailedness of the expert's analysis \cite{And:01, And:06}
of the non-public raw data.
From a point of view of general scientific
methodology, we find it however desirable that important results of
fundamental physics that require extensive
numerical treatment can be repeated by a broad public of non-expert scientists.
The preliminary analysis and the code given below is indeed intended for
those who like to get their own opinion in brief.
Furthermore, minor modifications
allow to test some alternative explanations the 
reader eventually may have in mind. A quick description for
getting started is found in section~\ref{sbs}. Though we cannot
give a detailed description of the program, some clarifying
comments are included in the quite self-explaining code (see
appendix).

\section{Methods}

\subsection{Limitations}

Our analysis is based on the reliability
of JPL's ephemeris files. As far as
Pioneer 10 and 11 is concerned, they contain an explicit warning:

\bq
`This trajectory is suitable for general historical purposes, but should
  be used cautiously for high precision or tracking data applications. This is
  due to potential dynamical mismatches between the Pioneer era models (DE-118,
  Lieske's E3 satellite theory of JUP035, SAT050, etc.) and the current modern
  solutions used by Horizons. For example, if the Pioneer 10 (11) solutions used
  here estimated planet or satellite ephemeris corrections at the time.
  However, the transformation from the original DE-118 planetary ephemeris
  coordinate system to the modern frame IS computed by Horizons. (...)
  The circumstances pertaining to the regeneration of the spacecraft
  trajectory source files are not well known.'
\eq

General remarks on the limitations of the accuracy of JPL ephemerides
are found in chap.~19 of the HORIZONS manual \cite{HORman}.

\subsection{General method}

We used the computer algebra system $Mathematica$ to calculate
spacecraft trajectories from known initial conditions and from the
positions of gravitating planetary bodies. The built-in-function
NDSolve with an explicit Runge-Kutta method was used to integrate
the equations of motion 
\be
\frac{d^2}{dt^2} \vec r(t) = - G \sum_i \frac{m_i (\vec r-\vec r_i)}{|\vec r-\vec r_i|^3}; 
\ \vec r(0)=\vec r_0; \  \frac{d}{dt}  \vec r(0)= \vec v_0,
\ee
while the sum is taken over all relevant bodies. Instead of calculating gravitational
forces using the gravitational constant $G$ and masses, the much
more accurate  Keplerian constants \cite{HORman}, p.~47, for the
sun and respective planets were implemented. For simplicity,
Mercury, Venus, Earth, Mars and the
asteroid belt masses were assumed to stay 
at sun's barycenter. Thus, for our approximate analysis
ephemerides of the outer planets and the sun were sufficient.
Since  the same $1/r^2$-law is obeyed, radiation pressure was
implemented by slightly diminishing the sun's effective mass (see code below for details)\fo{For large distances, 
the antenna can be assumed to be directed to the sun.}.
We estimated an effective surface\fo{Calculated from the diameter
of the antenna as given on p.~2 \cite{And:01}; The value given on p.~28 is different.} of the spacecraft of $5.9 m^2$ and an albedo of $0.7$.

\subsection{Data acquisition}

\begin{figure}[h]
\includegraphics[width=15cm]{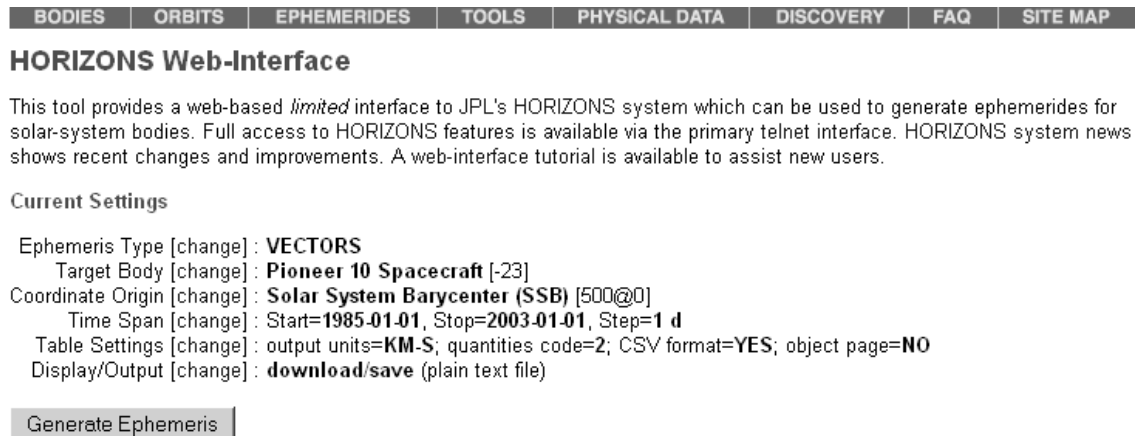}
\caption{Screenshot of HORIZONS site}
\label{screen}
\end{figure}

All the data needed were downloaded from JPL's ephemeris site HORIZONS which contains
spacecraft trajectories and planetary orbits. With our code, the reader can investigate
any period of interest. For our analysis, we downloaded the orbits of the sun, of Jupiter, Saturn,
Uranus and Neptune\fo{Other bodies may easily be included; download the respective ephemeris, 
add the name to variable planetnames, and set the variable plmax to the desired value.}
and the trajectories of Pioneer 10 and 11. We chose cartesian coordinates (setting VECTORS) with the solar system
barycenter as origin. A time step of 1 day was sufficient for our purposes\fo{Flyby modeling
would require much smaller steps and quadrupole moments of planets.}, though the code
works with smaller steps without changes. For the numerical treatment, data were 3-D-spline
interpolated. Due to lack of information, we did not do any maneuver modeling.

\subsection{Modeling}

The beginning of our analysis in 1987 coincides with the analysis carried out by \cite{And:01},
see also \cite{Lam:06}. The last signal from Pioneer 10 was received in 2002, while the RTG of Pioneer
11 became inoperable in Sept 1995. No reasonable analysis can be done after those dates;
we do not even know if
the very last signals have been included for the  generation of the trajectory files.
The time spans for the downloaded files can be chosen from 1985-2003 anyway, since
start and end can be chosen separately in $simulatespacecraft$[]. No spacecraft data
is available before the respective starts in 1972 and 1973. 

\section{Results - a preliminary analysis}

We compare predicted (simulated with conventional gravity) and observed (HORIZONS)
values for position, velocity, and acceleration (fig.~2
and 3). All quantities show a significant deviation.

\subsection{Comparison with position}
Here we consider the predicted $r(t)$ of our simulation with the position $r_{H}(t)$
in the ephemeris file\fo{While running $simulatespacecraft$[...], set option $vflag$ to $0$.}.
Velocities and accelerations are obtained by numerical differentiation. Though Pioneer 10 and 11
do not move precisely in the direction away from the solar system barycenter, for simplicity the radial
components only are analyzed.

\paragraph{An anomalous acceleration} is clearly visible for both Pioneer 10 and 11 (bottom graphs).
While the median
(minimizes absolute deviation) is $-7.60 \times 10^{-10} m s^{-2}$ and $-8.29 \times 10^{-10} m s^{-2}$
respectively, a quadratic fit to the acceleration data yields $-7.25 \times 10^{-10} m s^{-2}$
and $-7.05 \times 10^{-10} m s^{-2}$ (Pioneer 10 and 11).

\paragraph{The anomalous velocity} decrease is shown in the middle of fig.~(2)
and (3). A simple estimate $a=\frac{\Delta v}{\Delta t}$ from the velocity data
yields $-10.34 \times 10^{-10} m s^{-2}$ for Pioneer 10 and $-8.72 \times 10^{-10} m s^{-2}$
for Pioneer 11.

\paragraph{The deviation in radial position} is shown on top of fig.(~2)
and (3). In this case, the estimate $a=\frac{2 \Delta r}{(\Delta t)^2}$ yields
$-13.7 \times 10^{-10} m s^{-2}$ and $-9.52 \times 10^{-10} m s^{-2}$.

Of course, different time spans yield slightly different values, as the reader may easily
verify. All the above values are calculated automatically by our program.


\includegraphics[width=10cm]{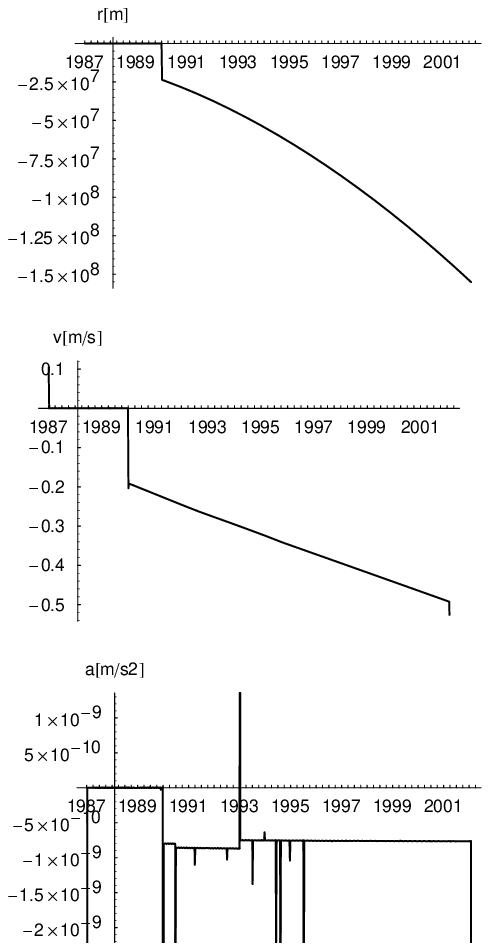}
\begin{center} Figure 2: Anomalous deviation (r), velocity (v) and acceleration (a) of Pioneer10 from 1987-2002.
\end{center}


\includegraphics[width=10cm]{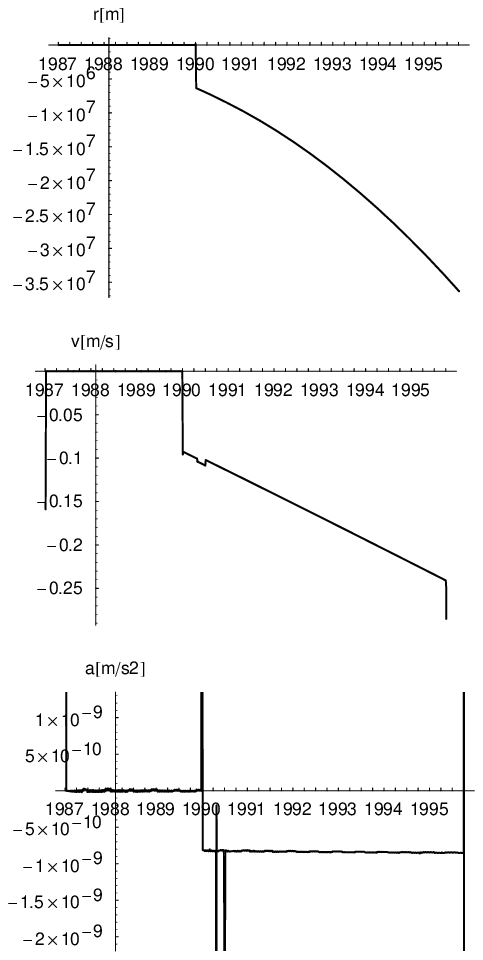}
\begin{center} Figure 3: Anomalous deviation (r), velocity (v) and acceleration (a) of Pioneer11 from 1987-1995.
\end{center}

\subsection{Comparison with velocities}
The derivatives $v(r)$ of our predicted $r(t)$ from the simulation are here compared to the velocities
$v_{H}(t)$ from HORIZONS\fo{While running $simulatespacecraft$[..], set option $vflag$ to $1$.}.
 Calculating $a$ from $\Delta v$ in this case yields
$-13.2 \times 10^{-10} m s^{-2}$ for Pioneer 10
and $-4.56 \times 10^{-10} m s^{-2}$ for Pioneer 11. The anomalous acceleration functions
(see bottom graphs of fig.~2 and~3) do practically not change. Run $simulatespacecraft$[-1,...,1] and
$generateplots$[-1] to get the respective plot. %

\section{Discussion}


Though the deviation of the observed quantities from the predicted ones are clearly visible
and confirm the order of magnitude of $a_p \approx c H_0$, there are a couple of results we cannot understand yet.

First of all, there is a big jump in the data on January 1st,
1990. Before that date, we cannot verify the known anomalous
acceleration at all, there seems to be an agreement with the predicted
trajectories. This must be due to a systematic error in
the ephemeris programs, i.e. a mismatch between data used then and
now, as mentioned above in the header of the JPL ephemeris file.

Further jumps in the acceleration of Pioneer 10, though of much
smaller amount, occurred on January 1st, 1993 and in June 1990.
The acceleration remains in the range of $a_p$, however. While
these jumps occur from a more or less constant level to another,
there are a couple of isolated disturbances, see bottom of
fig.~(2) and (3). Most likely those
disturbances are due to spacecraft maneuvers we did not model at
all. Taking the median, those disturbances are practically taken
out from the analysis. One should keep in mind however that even
the estimates from $\Delta r$ and $\Delta v$ (which were affected
by maneuvers) yielded $a_p$ in the correct order of magnitude.
Thus in no case the Pioneer anomaly can be an artifact of maneuver
mismodeling.

Looking at the acceleration plots at a high resolution we did not show here, a sinewave
disturbance with amplitude of about $10^{-11} m s^{-2}$ appears. The period however varies continuously
from about 12 days (Pioneer 11, around 1987) to 55 days (Pioneer 10, around 1997) and therefore
cannot be attributed to a mismodeling
of lunar ephemerides or a neglection 
of tidal effects
of the receiver station. It is clearly {\em not\/} a manifestation of the annual and diurnal signal
published in \cite{And:01} and must be due to some other systematic error.

\section{Conclusions}
Despite the limitations of JPL's ephemeris data, we could verify the anomalous
acceleration of Pioneer 10 and 11, i.e. $a_p$ cannot be due to a mismodeling of
gravitational attraction, maneuvers or radiation pressure. With our code, the reader
can easily verify or falsify further hypotheses on the origin of the anomaly. To account for
the effect of dust, Kuiper or asteroid belt masses, dark energy etc. it suffices to run the program
with different parameters or to add minor changes to the code.
We hope that this hands-on demonstration will develop further the interest and critical
acceptance of this outstanding effect.

\paragraph{Acknowledgement.} Though we are grateful for any comments, please 
understand that we cannot guarantee functionality or give further 
support for getting this program to run on your computer.

\section{Appendix:  data preparation and source code}

\subsection{Step-by step procedure in 15 minutes}
\label{sbs}
\begin{enumerate}
\i Create your directory `pioneer' and copy all of the following files in there
\i Goto NASA's ephemeris site HORIZONS: {\em http://ssd.jpl.nasa.gov/horizons.cgi \/}
\i Change ephemeris type to VECTORS (vector table)
\i Set Coordinate origin to [500@0], ecliptic and mean equinox of reference epoch ICRF/J.2000.0 
\i Set time span from 1985-01-01 to 2003-01-01, step 1 day
\i Chose table settings $km, km/s$, quantities code $=2$ (two-state vector $x,y,z,vx,vy,vz$), CSV format $= YES$, object page $=NO$.
\i Display/Output: download/save as plain text file.
\i Chose target body: sun  (sol)
\i Generate ephemeris and save as {\em sun-85-03.txt\/}. See also screenshot fig.~\ref{screen}.
\i Proceed in the same manner with Jupiter, Saturn, Uranus, Neptune, Pioneer 10 and 11
as target bodies.
\i Cross-check if the file ending for the Pioneers corresponds to the last two entries of `spans' (line 2 of the code).
\i Chose `barycenter' where appropriate (otherwise you'll miss Jupiter's satellites) and
 save as  {\em neptune-85-03.txt\/} etc.
\i Type source from appendix or download it from {\em www.alexander-unzicker.de/pioneer.txt\/}\fo{
Do not paste and copy from LaTeX source code, this will create error messages.}
\i Change in the first line the path to your pioneer directory
\i Open a Mathematica *.nb file and run the following commands (see also file end of source):
 \bq
 SetDirectory["c:$\setminus \setminus$yourpioneerdirectory"]; 
 (* insert a $\setminus \setminus$ for any $\setminus$   in the path*) \\
  $<<$pioneer.txt; \\
  planetini[5,"-85-03"];\\
 simulatespacecraft[-1, 46800.5(*day start 5.1.87*), 8.75(*insert years*), 0 (*option vflag*)];
 genererateplots[-1];
 \eq
\i Proceed likewise with simulatespacecraft[-2, 46798.5(* 3.1.87*), 15.08,0] for Pioneer 10.
\end{enumerate}

\subsection{Source code}
\mbox{(*************** global settings *****************)} \\
\mbox{planetnames = \{"sun", (*"merkur", "venus","earth","mond","mars",*)"jupiter", "saturn","uranus",} \\
\mbox{"neptune","pluto","pioneer10","pioneer11"\};} \\
\mbox{spans=\{"-85-03","-85-03"\}; (*fileendings of spacecraft emphemerides*)} \\
\mbox{mjd=2400000; (** mean julianian day correction**)} \\
\mbox{day=3600*24; (* SI unit is seconds *)} \\
\mbox{(************** constants for radiation pressure modeling *******)} \\
\mbox{cc = 299792458;  (*speed of light*)  AU = 1.496*10\symbol{94}(11); (* astronomic unit *)} \\
\mbox{albedo=0.7; solarK=1367*AU\symbol{94}2/cc*(1+albedo); } \\
\mbox{surfaceP=\{5.9,5.9\}; (* effective surfaces for radiation pressure in m\symbol{94}2 *)} \\
\mbox{massP=\{258,259\};    (* spacecraft masses*)} \\
\mbox{(*Instead of masses, the much more accurate Kepler constants are used, see HORIZONS manual p.47 ****)} \\
\mbox{KKs=10\symbol{94}9\{132712440017.98698,126712767.857796, 37940626.061137281,5794549.00707,} \\
\mbox{6836534.0638792608,981.6008877\}; (*kepler's constants**)} \\
\mbox{(** all masses inside the asteroid belt (3 10\symbol{94}(21) kg ) added to sun (KKs[[1]]) **)} \\
\mbox{KKs[[1]]+=10\symbol{94}9*Apply[Plus, \{22032.080486417923, 324858.59882645978,398600.43289693922,} \\
\mbox{4902.8005821477636, 42828.314258067119, 200 (*asteroid belt estimate*)\}];} \\
\mbox{(** for plots: display year numbers at axes instead of days *)} \\
\mbox{subti=4; ti87 =Transpose[\{Table[46796.5+365.25i/subti,\{i,0,16subti\}],} \\
\mbox{Table[If[IntegerQ[i/subti]==True,1987+i/subti,""],\{i,0,16 subti\}]\}]; } \\
\mbox{ti88 =Transpose[\{Table[46796.5+365.25i/subti,\{i,0,16 subti\}],} \\
\mbox{Table[If[IntegerQ[i/subti/2]==True,1987+i/subti,""],\{i,0,16 subti\}]\}]; } \\
\mbox{SCplots=Table[\{0,0,0\}, \{i,10\}]; (* plot variable*)} \\
\mbox{\$DefaultFont = \{"Arial", 8\};} \\
\mbox{readplanet[name\_,ending\_]:=Block[\{qwe, wer\}, filename=planetnames[[name]] $<$ $>$ending $<$ $>$".txt";} \\
\mbox{If[FileInformation[filename]==\{\},Print["Missing ephemeris file. Stop."];Break[]];} \\
\mbox{IO=3; (* interpolation order of splines *)} \\
\mbox{qwe = Import[filename, "CSV"];} \\
\mbox{cut1 = Position[qwe, "\$\$SOE"][[1, 1]]; cut2 = Position[qwe, "\$\$EOE"][[1, 1]];} \\
\mbox{wer = Take[Drop[qwe, cut1], cut2 - cut1 - 1];} \\
\mbox{xyz = 1000*Transpose[Take[Transpose[wer], \{3, 5\}]]; (*** km - m  factor ***)} \\
\mbox{vxyz = 1000*Transpose[Take[Transpose[wer], \{6, 8\}]];} \\
\mbox{time = Flatten[Transpose[wer][[1]]] - mjd;} \\
\mbox{x1=Interpolation[Transpose[\{time,Transpose[xyz][[1]]\}],InterpolationOrder $->$ IO];} \\
\mbox{y1=Interpolation[Transpose[\{time,Transpose[xyz][[2]]\}],InterpolationOrder $->$ IO];} \\
\mbox{z1=Interpolation[Transpose[\{time,Transpose[xyz][[3]]\}],InterpolationOrder $->$ IO];} \\
\mbox{(** x,y,z, as function, interpolated***)} \\
\mbox{vx1=Interpolation[Transpose[\{time,day*Transpose[vxyz][[1]]\}],InterpolationOrder $->$ IO];} \\
\mbox{vy1=Interpolation[Transpose[\{time,day*Transpose[vxyz][[2]]\}],InterpolationOrder $->$ IO];} \\
\mbox{vz1=Interpolation[Transpose[\{time,day*Transpose[vxyz][[3]]\}],InterpolationOrder $->$ IO];} \\
\mbox{];} \\
\mbox{(* simple differentiation routine for interpolating functions of different range, step: 1 day*)} \\
\mbox{numdiff[tab\_,st\_(*start time in JD*),en\_(*end time in JD*)]:=Block[\{tab1p,tab1m,tab2m, tab2p\},} \\
\mbox{tab1p = RotateRight[tab, 1]; tab1m = RotateRight[tab, -1];} \\
\mbox{dtab = Flatten[\{tab[[2]]-tab[[1]],Drop[Drop[(tab1m - tab1p)/2, \{1\}], \{-1\}],tab[[-1]]-tab[[-2]]\}/day];} \\
\mbox{ttime = Table[i, \{i, st, en, 1\}];} \\
\mbox{dtabtime = Transpose[\{ttime, dtab\}];} \\
\mbox{dlist=Interpolation[dtabtime];dlist];} \\
\mbox{planetini[plmax\_,plfileending\_]:=Block[\{\},} \\
\mbox{(***** Reading data of all relevant planets 1 sun 2 jup, 3 sat, 4 Ura, 5 nep *********)} \\
\mbox{xx=yy=zz=\{\}; } \\
\mbox{Print["Initialize planets..."];} \\
\mbox{For[k=1,k $<$ =plmax,k++,readplanet[k,plfileending];AppendTo[xx,x1];AppendTo[yy,y1];AppendTo[zz,z1]]];} \\
\mbox{simulatespacecraft[craftflag\_,start\_,yrs\_,vflag\_(* set to 1 for comparing observed velocities*)]:=Block[\{\},} \\
\mbox{(********** starting values **************)} \\
\mbox{(** -1 for pio 11, -2, for pio 10,  anything else runs automatically*)} \\
\mbox{fileending=spans[[craftflag]];} \\
\mbox{sta=start;} \\
\mbox{(* radiation pressure modelled as missing solar mass *)} \\
\mbox{KKs[[1]]-=solarK*surfaceP[[craftflag]]/massP[[craftflag]];} \\
\mbox{end=Ceiling[start+yrs*365.25]+.5;} \\
\mbox{readplanet[craftflag, fileending]; } \\
\mbox{(**** starting values for simulation ***)} \\
\mbox{\{x0,y0,z0\}=\{x1[t],y1[t],z1[t]\} /. t$->$start; } \\
\mbox{\{vx0,vy0,vz0\}=\{vx1[t],vy1[t],vz1[t]\} /. t$->$start; (* new*)} \\
\mbox{Print["computing trajectory for ", planetnames[[craftflag]], " ... "];} \\
\mbox{(***** numeric integration of the equations of motion by Runge-Kutta****)} \\
\mbox{plmax=Length[zz];} \\
\mbox{loe = NDSolve[\{} \\
\mbox{x''[t]==Apply[Plus,Table[-day\symbol{94}2 KKs[[i]](x[t]-xx[[i]][t])/((x[t]-xx[[i]][t])\symbol{94}2+} \\
\mbox{(y[t]-yy[[i]][t])\symbol{94}2+(z[t]-zz[[i]][t])\symbol{94}2)\symbol{94}(3/2), \{i, 1, plmax\}]], } \\
\mbox{y''[t]==Apply[Plus, Table[-day\symbol{94}2 KKs[[i]](y[t]-yy[[i]][t])/((x[t]-xx[[i]][t])\symbol{94}2+} \\
\mbox{(y[t]-yy[[i]][t])\symbol{94}2+(z[t]-zz[[i]][t])\symbol{94}2)\symbol{94}(3/2),\{i,1, plmax\}]],} \\
\mbox{z''[t]==Apply[Plus,Table[-day\symbol{94}2 KKs[[i]](z[t]-zz[[i]][t])/((x[t]-xx[[i]][t])\symbol{94}2+} \\
\mbox{(y[t]-yy[[i]][t])\symbol{94}2+(z[t]-zz[[i]][t])\symbol{94}2)\symbol{94}(3/2), \{i, 1, plmax\}]],} \\
\mbox{x[start]==x0,y[start]==y0,z[start]==z0,} \\
\mbox{x'[start]==vx0,y'[start]==vy0, z'[start]==vz0\},} \\
\mbox{\{x[t], y[t],z[t]\}, \{t, start, end\},Method$->$"ExplicitRungeKutta"];} \\
\mbox{r[t\_] := Sqrt[loe[[1, 1, 2]]\symbol{94}2 + loe[[1, 2, 2]]\symbol{94}2 + loe[[1, 3, 2]]\symbol{94}2]; (* solution r(t) *)} \\
\mbox{v[t\_]:=D[r[t],t]/day;} \\
\mbox{a[t\_]:=D[v[t],t]/day;} \\
\mbox{readplanet[craftflag, fileending];  (* reading spacecraft data again*)} \\
\mbox{radius[t\_]:= Sqrt[x1[t]\symbol{94}2 + y1[t]\symbol{94}2 + z1[t]\symbol{94}2];      (* observed r(t) from Horizons*)} \\
\mbox{vradial[t\_]:= (vx1[t]*x1[t] + vy1[t]*y1[t] + vz1[t]*z1[t])/day/radius[t]; (*radial component*)} \\
\mbox{vobserv=Interpolation[Transpose[\{Table[i,\{i,start,end,1\}],Table[vradial[t],\{t,start,end,1\}]\}]];} \\
\mbox{ranom[t\_]:=radius[t]-r[t]; (* anomaly  of position **)} \\
\mbox{(**** vflag=1 compares to observed velocity, vflag=0, to position***)} \\
\mbox{If[vflag==1, vvv=numdiff[Table[r[t], \{t, start, end, 1\}], start, end];} \\
\mbox{  Clear[vanom];vanom[t\_]:=vobserv[t]-vvv[t],vanom=numdiff[Table[ranom[t], \{t, start, end, 1\}],start,end]];} \\
\mbox{aanom=numdiff[Table[vanom[t], \{t, start, end, 1\}],start,end];} \\
\mbox{ap=((vanom[t]/.t $->$end)-(vanom[t]/.t $->$start))/(end-start)/day;  (* ap estimate from velocities *)} \\
\mbox{ap2=2((ranom[t]/.t $->$end)-(ranom[t]/.t $->$start))/(end-start)\symbol{94}2/day\symbol{94}2; (* ap from position ***)} \\
\mbox{Print["anomalous acceleration (from v and r):  ", ap, "   ", ap2];} \\
\mbox{atab=Table[aanom[t], \{t, sta, end\}];} \\
\mbox{median=Median[atab];} \\
\mbox{abwei=FindMinimum[Sum[Abs[x - atab[[i]]], \{i, Length[atab]\}], \{x, 0\}][[2, 1, 2]];} \\
\mbox{abwei2=FindMinimum[Sqrt[Sum[Abs[x - atab[[i]]]\symbol{94}2, \{i, Length[atab]\}]], \{x, 0\}][[2, 1, 2]];} \\
\mbox{Print["Median, absolute, quadratic deviation minimized:  ", median, "   ", abwei, "   ", abwei2];} \\
\mbox{KKs[[1]]+=solarK*surfaceP[[craftflag]]/massP[[craftflag]]]; (*take out again radiation from solar mass*)} \\
\mbox{generateplots[body\_]:=Block[\{\},AO=sta+400;(* clean plot: axes origin shifted days in the future *)} \\
\mbox{tic=If[body==-1,ti87,ti88];} \\
\mbox{SCplots[[body,1]]=\{Plot[radius[t]-r[t],\{t, sta, end\}, Ticks$->$\{tic, Automatic\},} \\
\mbox{AxesOrigin$->$\{AO, 0\},AxesLabel $->$\{"", "r"\},PlotRange$->$All]]\};} \\
\mbox{SCplots[[body,2]]=\{Plot[vanom[t],\{t, sta, end\}, Ticks$->$\{tic, Automatic\},} \\
\mbox{AxesOrigin$->$\{AO, 0\},AxesLabel $->$\{"", "v"\},PlotRange$->$All]\};} \\
\mbox{SCplots[[body,3]]=\{Plot[aanom[t],\{t, sta, end\}, Ticks$->$\{tic, Automatic\},} \\
\mbox{AxesOrigin$->$\{AO, 0\},AxesLabel $->$\{"", "a"\}]\}];} \\

\end{document}